\documentclass[letterpaper, 11pt]{article}
\usepackage{amsfonts, amsmath, amssymb, amsthm, graphicx, cite}

\newtheorem{theorem}{Theorem}
\newtheorem{lemma}{Lemma}

\newcommand{\Fto}[3]{\setlength{\arraycolsep}{0.0em}{}_2F_1\bigg(\begin{array}{c}{#1}\\{#2}\end{array};{#3}\bigg)\setlength{\arraycolsep}{5 pt}}
\newcommand{\Ftt}[3]{\setlength{\arraycolsep}{0.0em}{}_3F_2\bigg(\begin{array}{c}{#1}\\{#2}\end{array};{#3}\bigg)\setlength{\arraycolsep}{5 pt}}
\newcommand{\Fpq}[3]{\setlength{\arraycolsep}{0.0em}{}_pF_q\bigg(\begin{array}{c}{#1}\\{#2}\end{array};{#3}\bigg)\setlength{\arraycolsep}{5 pt}}

\newcommand{\SNR}{\mathrm{SNR}}

\newcommand{\ud}{\mathrm{d}}

\mark{{}{}}

\date{}

\begin{document}

\title{On the Capacity and Mutual Information of Memoryless Noncoherent Rayleigh-Fading Channels}

\author{S\'ebastien de la Kethulle de Ryhove\thanks{S.~de la Kethulle de Ryhove and G.~E.~\O ien are with the Department of Electronics and Telecommunications, 
Norwegian University of Science and Technology, N--7491 Trondheim, Norway (e-mails: \{delaketh, oien\}@iet.ntnu.no).}, 
Ninoslav Marina\thanks{N.~Marina was with the Swiss Federal Institute of Technology (EPFL), CH--1015 Lausanne, Switzerland, and is currently with
Sowoon Technologies, PSE Technology Park, CH--1015 Lausanne, Switzerland \mbox{(e-mail:} ninoslav.marina@sowoon.com).}, and Geir E. \O ien$^*$}

\maketitle

\begin{abstract}
The memoryless noncoherent single-input single-output (SISO) Rayleigh-fading channel is considered. Closed-form expressions for the mutual information between
the output and the input of this channel when the input magnitude distribution is discrete and restricted to having two mass points are derived,
and it is subsequently shown how these expressions can be used to obtain closed-form expressions for the capacity of this channel
for signal to noise ratio (SNR) values of up to approximately $0 \; \mathrm{dB}$, and a tight capacity lower bound for SNR values
between $0 \; \mathrm{dB}$ and $10 \; \mathrm{dB}$.
The expressions for the channel capacity and its lower bound are given as functions of a parameter which can be obtained via numerical
root-finding algorithms. 
\end{abstract}


\section{Introduction}

Wireless communication channels in which neither the transmitter nor the receiver possess any knowledge of the channel propagation coefficients
(also known as noncoherent channels) have recently been receiving a considerable amount of
attention~\cite{marzetta_hochwald, zheng_tse, liang, abou-faycal, taricco_elia, lapidoth}. Such channels arise whenever the channel
coherence time is too short to obtain a reliable estimate of the propagation coefficients via the standard pilot symbol technique (high mobility
wireless systems are a typical example of such a scenario). They are currently less well understood than coherent channels, in which the channel state
is assumed to be known to the receiver (and sometimes also the transmitter). 

In this correspondence, we consider the memoryless noncoherent single-input single-output (SISO) Rayleigh-fading
channel, which was studied under the assumption of an average power constrained input in e.g.~\cite{abou-faycal, taricco_elia, lapidoth}. 
In~\cite{abou-faycal}, Abou-Faycal \textit{et al.} rigorously proved (in the average power constrained input case)
that the magnitude of the capacity-achieving distribution is discrete with a finite number of mass points, one
of these mass points being necessarily located at the origin (zero magnitude). Using numerical optimisation algorithms, the authors
also empirically found that a magnitude distribution with two mass points achieves capacity at low signal to noise ratio (SNR) values,
and that the required number of mass points to achieve capacity increases monotonically with the SNR. 
Numerical optimisation algorithms remain however the only way to find the number
of mass points of the capacity-achieving magnitude distribution for a given SNR.

Another important reference on the average power constrained memoryless noncoherent SISO Rayleigh fading channel is
the work of Taricco and Elia~\cite{taricco_elia}, where lower and upper capacity bounds were established, and it was also proved that for high SNR values
the capacity only grows double-logarithmically in the SNR. The upper bound from~\cite{taricco_elia} was subsequently tightened by Lapidoth and
Moser~\cite{lapidoth} in the framework of a more general study on capacity bounds and multiple-antenna systems on flat-fading channels.

The problem of finding the capacity-achieving magnitude distribution of the memoryless noncoherent SISO Rayleigh-fading channel in the
average power constrained input case can be solved for low SNR values by using numerical optimisation
algorithms as in~\cite{abou-faycal}, but the optimisation problem becomes intractable for high SNR values due to its high dimensionality.
An additional difficulty arises due to the fact that closed-form solutions for the integrals
appearing in the expression for the channel mutual information when the input magnitude distribution is discrete are not available in the literature.
Consequently, numerical integration algorithms must be repeatedly used when performing such an optimisation.

In this letter, we derive closed-form expressions for the mutual information of the noncoherent Rayleigh-fading
channel when the input magnitude distribution is discrete and has two mass points, thus completely eliminating the need for numerical integration in order
to compute the mutual information in such a case. For low SNR values (of up to approximately $0 \; \mathrm{dB}$)
and an average power constrained input, in which case the capacity-achieving magnitude distribution is discrete with two mass points,
these closed-form expressions additionally enable us to write the channel capacity as a function of a single parameter which can be obtained via
numerical root-finding techniques. This capacity expression also becomes a tight capacity lower bound when the SNR takes values
between $0\; \mathrm{dB}$ and $10\;\mathrm{dB}$.

Part of the material in this correspondence can also be found in~\cite{marina_springer, marina_phd}, where an expression for the mutual information of the
noncoherent Rayleigh-fading channel, when the input magnitude distribution is discrete and restricted to having only two mass points, was presented in the framework of a
study regarding the capacity region of a two-user multiple-access channel in which the channel state is known neither to the transmitters nor to the receiver.
The most important additional contributions of this letter lie firstly in a fully detailed and rigorous proof of the validity of this expression,
secondly in the derivation of alternative analytical expressions for the same quantity (which naturally follow from the structure of our
proof), thirdly in the discussion of a special case in which the expression provided in~\cite{marina_springer, marina_phd} can be simplified, fourthly in noting that
the hypergeometric functions~\cite{table} appearing in the expression from~\cite{marina_springer, marina_phd} can also be expressed in terms of the incomplete
beta function~\cite{table}, and fifthly in showing how to obtain analytical expressions for the derivative of the mutual information of the noncoherent
Rayleigh-fading channel (when the input magnitude distribution is discrete with two mass points) with respect to
parameters of interest, with applications to capacity calculations and the derivation of a capacity lower bound.

The remainder of this letter is organised as follows: the channel model is introduced in Sec.~\ref{sec:cm}, closed-form
expressions for the mutual information when the channel input magnitude distribution is discrete and has two mass points being subsequently derived in
Sec.~\ref{sec:closedformI}. Applications to capacity calculations (for SNR values of up to approximately $0 \; \mathrm{dB}$) and the derivation of
a capacity lower bound (for SNR values between $0\; \mathrm{dB}$ and $10\;\mathrm{dB}$) in the average power constrained input case are then discussed
in Sec.~\ref{sec:cap}, and conclusions finally drawn in Sec.~\ref{sec:conclusions}.

\section{Channel Model}
\label{sec:cm}

We consider discrete-time memoryless noncoherent SISO Rayleigh fading channels of the form~\cite{abou-faycal}
\begin{equation}
  \label{eq:fading_channel}
  V_k = A_k U_k + W_k,
\end{equation}where for each time instant $k \in \mathbb{N}$, $A_k \in \mathbb{C}$, $U_k \in \mathbb{C}$, 
$V_k \in \mathbb{C}$, and $W_k \in \mathbb{C}$ respectively represent the channel fading coefficient, the transmitted symbol, the received symbol, and
the channel noise. The elements of the sequences $\{A_k\}$ and $\{W_k\}$ are assumed to be zero mean i.i.d. circularly symmetric complex Gaussian random variables
with variances respectively equal to $1$ and $\sigma^2 > 0$, i.e. $A_k \sim \mathcal{N}_{\mathbb{C}}(0, 1)$ and $W_k \sim \mathcal{N}_{\mathbb{C}}(0, \sigma^2)$.
It is also assumed that the elements of the sequences $\{A_k\}$ and $\{W_k\}$ are mutually independent. The channel~(\ref{eq:fading_channel}) being stationary
and memoryless, we henceforth omit the notation of the time index $k$.

$A$ and $W$ being circularly symmetric complex Gaussian distributed, it follows that conditioned on a value $u$ of $U$, the channel output $V$
also is circularly symmetric complex Gaussian distributed, with mean value zero and variance $|u|^2 + \sigma^2$. Consequently, conditioned on the input $U$, 
the channel output $V$ is distributed according to the law
\begin{equation}
  \label{eq:fVU}
  f_{V|U}(v|u) = \frac{1}{\pi(|u|^2+\sigma^2)} \; e^{\frac{-|v|^2}{|u|^2+\sigma^2}}.
\end{equation}
Note that since $f_{V|U}(v|u)$ is independent of the phase $\arg u$ of the input signal, the latter quantity cannot carry any information.
If we now make the variable transformations
\begin{equation}
  \setlength{\arraycolsep}{0.0em}\left\{\begin{array}{rcl}
  X &{}={}& |U|\\ 
  Y &{}={}& |V|
  \end{array}\right.,\setlength{\arraycolsep}{5pt} \qquad \qquad X, Y \geq 0,
\end{equation}
we obtain an equivalent channel, the probability density of the output $Y$ of which, conditioned on a value $x$ of its input $X$, is given by
\begin{equation}
  \label{eq:fYX}
  f_{Y|X}(y|x) = \frac{2y}{x^2+\sigma^2} \; e^{\frac{-y^2}{x^2+\sigma^2}} \qquad X, Y \geq 0.
\end{equation}
Let $I(X;Y)$ denote the mutual information~\cite{cover} between the input and output of the channel
with transition probability~(\ref{eq:fYX}), and let $f_X(x)$ and $f_Y(y)$ respectively denote the channel input and output probability density functions.
Note that since $f_{V|U}(v|u)$ is independent of $\arg u$ and $\arg v$ we always have $I(X;Y) = I(U;V)$,
where $I(U;V)$ denotes the mutual information between the input and output of the channel with transition probability~(\ref{eq:fVU}).

\section{Closed-Form Expressions For $I(X;Y)$ When The Input Distribution Has Two Mass Points}
\label{sec:closedformI}

For a given input probability density function $f_X(x)$, the mutual information $I(X;Y)$ of the channel with transition probability~(\ref{eq:fYX}) is
given by\footnote{In this letter, logarithms will be taken in base $e$, and mutual information will be measured in nats.}
\begin{equation}
  \label{eq:IXY}
  I(X;Y) = \int_0^\infty \int_0^\infty f_{Y|X}(y|x) f_X(x) \log \frac{f_{Y|X}(y|x)}{f_Y(y)} \, \ud y \, \ud x,
\end{equation}
with $f_Y(y) = \int_0^\infty f_{Y|X}(y|x) f_X(x) \, \ud x$. In this section, we derive a closed-form expression for $I(X;Y)$ when the input
probability density function $f_X(x)$ has the form
\begin{equation}
  \label{eq:inputfx}
  f_X(x) = a_1 \delta(x - x_1) + a_2 \delta(x - x_2),
\end{equation}
where $\delta(\cdot)$ denotes the Dirac distribution, $a_1$ and $a_2$ are constants which, $f_X(x)$ being a probability density function,
must be such that $0 \leq a_1, a_2 \leq 1$ and $a_1 + a_2 = 1$, $x_1, x_2 \geq 0$ by virtue of the fact that $X = |U| \geq 0$, and
we assume without loss of generality that $x_1 \leq x_2$.

We will in addition assume that $0 < a_1, a_2 < 1$, that $x_1 \neq x_2$, and that $x_1 = 0$. The reason for the first two assumptions is that
the case where either $a_1 = 0$, $a_1 = 1$, or $x_1 = x_2$ is of little interest since $f_X(x)$ then reduces to a probability density function with only one mass point
and the mutual information vanishes.
The reason for assuming that $x_1 = 0$ is that -- when the input probability distribution must satisfy an average power
constraint of the form $E\left[X^2\right] \leq P$ for a given power budget $P$ -- the capacity-achieving input distribution of
the channel with transition probability $f_{Y|X}(y|x)$
given in~(\ref{eq:fYX}) always has one mass point located at the origin~\cite{abou-faycal}, and consequently the case $x_1 = 0$ is one of great practical importance.
We however would like to point out that the case $x_1 \neq 0$ can be treated in exactly the same manner as below and that we choose not
do so in order to keep all equations as simple as possible. 

With this choice for $f_X(x)$, the mutual information~(\ref{eq:IXY}) becomes
\begin{equation}
  \label{eq:mutualI}
  I(X;Y) = -a_1 - a_1 \log \sigma^2 - a_2 - a_2 \log(x_2^2 + \sigma^2) 
           - \sum_{k=1}^2 \int_0^\infty \hspace{-1.2 pt} \frac{2y \, a_k}{x_k^2 + \sigma^2} e^{\frac{-y^2}{x_k^2 + \sigma^2}} 
           \log \left( \sum_{l=1}^2 \frac{a_l}{x_l^2 + \sigma^2} \, e^{\frac{-y^2}{x_l^2 + \sigma^2}} \right) \ud y.
\end{equation}
The difficulty to find a closed-form expression for $I(X;Y)$ lies in finding an expression for integrals of the form (with $x \in \{0, x_2\}$)
\begin{equation}
  \label{eq:Jx}
  J(x) \triangleq \hspace{-3.5 pt} \int_0^\infty \hspace{-3.5 pt} \frac{2y}{x^2 + \sigma^2} e^{\frac{-y^2}{x^2 + \sigma^2}} 
                     \log \left( \frac{a_1}{\sigma^2} e^{\frac{-y^2}{\sigma^2}} + \frac{a_2}{x_2^2 + \sigma^2} e^{\frac{-y^2}{x_2^2 + \sigma^2}} \right) \ud y,
\end{equation}
which appear in~(\ref{eq:mutualI}). Integrals resembling $J(x)$ do not appear in tables such as~\cite{table}, and to the best of the
authors' knowledge no closed-form expression is currently available in the literature. How to derive such closed-form expressions for $J(x)$,
which can then be used to obtain closed-form expressions for the
mutual information $I(X;Y)$ when the input probability density function $f_X(x)$ is of the form~(\ref{eq:inputfx}), is shown in Sec.~\ref{sec:Jxclosed} below.

\subsection{Closed-Form Expressions for $J(x)$}
\label{sec:Jxclosed}

Let us define 
\begin{equation}
   \label{eq:alphabeta}
   \alpha \triangleq \frac{x_2^2}{x_2^2 + \sigma^2} \frac{x^2 + \sigma^2}{\sigma^2}, \qquad \beta \triangleq \frac{a_2}{a_1} \frac{\sigma^2}{x_2^2 + \sigma^2},
\end{equation}
which remembering the above assumptions can be seen to be always strictly positive, and let us in addition introduce the strictly positive quantity
\begin{equation}
    \label{eq:ys2}
    y_*^2 \triangleq - \frac{\sigma^2 (x_2^2 + \sigma^2)}{x_2^2} \, \log \beta 
\end{equation}
whenever $\log \beta < 0$ (i.e. when $0 < \beta < 1$). Note that when this is the case, we have
\begin{equation}
  \frac{a_1}{\sigma^2} e^{\frac{-y_*^2}{\sigma^2}} =\frac{a_2}{x_2^2 + \sigma^2} e^{\frac{-y_*^2}{x_2^2 + \sigma^2}}
\end{equation}
for the $y_*^2 > 0$ defined in~(\ref{eq:ys2}). We now provide expressions for $J(x)$ in three different cases.

\subsubsection{Case I:  $\alpha \in \{1, \frac{1}{2}, \frac{1}{3}, \ldots \}$}
\label{sec:caseI}

Noting that 
\begin{equation}
  \label{eq:logid}
  \log \left( \frac{a_1}{\sigma^2} e^{\frac{-y^2}{\sigma^2}} + \frac{a_2}{x_2^2 + \sigma^2} e^{\frac{-y^2}{x_2^2 + \sigma^2}} \right) 
       = \frac{-y^2}{x_2^2+\sigma^2} + \log \frac{a_2}{x_2^2 + \sigma^2} + \log\left( 1 + \beta^{-1} e^{\frac{-\alpha y^2}{x^2+\sigma^2}}\right),
\end{equation}
it is easy to see that $J(x) = J_{11}(x) + J_{12}(x) + J_{13}(x)$, with
\setlength{\arraycolsep}{0.0em}\begin{eqnarray}
  \label{eq:J11}
  J_{11}(x) &{}={}& \int_0^\infty \frac{-2y^3}{(x^2 + \sigma^2)(x_2^2 + \sigma^2)}e^{\frac{-y^2}{x^2 + \sigma^2}} \ud y = -\frac{x^2 + \sigma^2}{x_2^2 + \sigma^2}, \\
  \label{eq:J12}
  J_{12}(x) &{}={}& \log \frac{a_2}{x_2^2 + \sigma^2} \int_0^\infty \hspace{-4 pt} 
                         \frac{2y}{x^2 + \sigma^2}e^{\frac{-y^2}{x^2 + \sigma^2}} \ud y = \log \frac{a_2}{x_2^2 + \sigma^2}, \qquad
\end{eqnarray}\setlength{\arraycolsep}{5pt}and
\begin{equation}
\label{eq:J13}  J_{13}(x) = \int_0^\infty \frac{2y}{x^2 + \sigma^2}e^{\frac{-y^2}{x^2 + \sigma^2}} 
                           \log\left( 1 + \beta^{-1} e^{\frac{-\alpha y^2}{x^2+\sigma^2}}\right) 
      = \left( 1-(-\beta)^{-\frac{1}{\alpha}}\right) \log \left( 1 + \beta^{-1}\right) - \sum_{k=1}^{1/\alpha} \frac{1}{k (-\beta)^{k-\frac{1}{\alpha}}}.
\end{equation}
The last equality is a direct consequence of the fact that, for any $a, b > 0$ we have, omitting the integration constant, 
\setlength{\arraycolsep}{0.0em}\begin{eqnarray}
                 \int \frac{2y}{a}\;e^{-y^2 \hspace{-1 pt}/a} \log\left( 1 + b \, e^{{-\alpha y^2}/{a}}\right) \ud y \hspace{9 cm} \nonumber \\
  \label{eq:indef1}      = \sum_{k=1}^{1/\alpha} \frac{1}{k} (-b)^{k-\frac{1}{\alpha}} e^{-k \alpha y^2 / a} 
                            - \Big(e^{-y^2 \hspace{-1 pt}/a} +(-b)^{-\frac{1}{\alpha}} \log b \Big)\log\left( 1 + b \, e^{{-\alpha y^2}/{a}}\right), \qquad 
\end{eqnarray}\setlength{\arraycolsep}{5pt}which can be verified by differentiating the expression on the right-hand side of the equality
sign with respect to $y$ and remembering that for any $q \neq 1$ and any positive integer $n$, we have $\sum_{k=1}^n q^k = (q-q^{n+1})/(1-q)$. 

\subsubsection{Case II:  $0 < \protect\beta < 1$ and $\protect\alpha \protect\notin \protect\{1, \protect\frac{1}{2}, \protect\frac{1}{3}, \protect\ldots \protect\}$}
\label{sec:caseII}

In this case, we have
$J(x) = J_{21}(x) + J_{22}(x) + J_{23}(x) + J_{24}(x) + J_{25}(x) + J_{26}(x)$, with
\setlength{\arraycolsep}{0.0em}\begin{eqnarray}
  \label{eq:J21} J_{21}(x) &{}={}& \int_0^{y_*} \frac{-2y^3}{\sigma^2(x^2 + \sigma^2)}e^{\frac{-y^2}{x^2 + \sigma^2}} \ud y, \\
  \label{eq:J22} J_{22}(x) &{}={}& \log \frac{a_1}{\sigma^2} \int_0^{y_*} \frac{2y}{x^2 + \sigma^2}e^{\frac{-y^2}{x^2 + \sigma^2}} \ud y, \\
  \label{eq:J23} J_{23}(x) &{}={}& \int_0^{y_*} \frac{2y}{x^2 + \sigma^2}e^{\frac{-y^2}{x^2 + \sigma^2}} 
                                  \log\left( 1 + \beta e^{\frac{\alpha y^2}{x^2+\sigma^2}}\right) \ud y, \\
  \label{eq:J24} J_{24}(x) &{}={}& \int_{y_*}^\infty \frac{-2y^3}{(x^2 + \sigma^2)(x_2^2 + \sigma^2)}e^{\frac{-y^2}{x^2 + \sigma^2}} \ud y, \\
  \label{eq:J25} J_{25}(x) &{}={}& \log \frac{a_2}{x_2^2 + \sigma^2} \int_{y_*}^\infty \frac{2y}{x^2 + \sigma^2}e^{\frac{-y^2}{x^2 + \sigma^2}} \ud y,
\end{eqnarray}\setlength{\arraycolsep}{5pt}and
\setlength{\arraycolsep}{0.0em}\begin{eqnarray}
  \label{eq:J26} J_{26}(x) &{}={}& \int_{y_*}^\infty \frac{2y}{x^2 + \sigma^2}e^{\frac{-y^2}{x^2 + \sigma^2}} 
                    \log\left( 1 + \beta^{-1} e^{\frac{-\alpha y^2}{x^2+\sigma^2}}\right) \ud y. \quad
\end{eqnarray}\setlength{\arraycolsep}{5pt}This can be seen by remembering~(\ref{eq:logid}), observing that we furthermore also have
\begin{equation}
  \label{eq:logid2}
  \log \left( \frac{a_1}{\sigma^2} e^{\frac{-y^2}{\sigma^2}} + \frac{a_2}{x_2^2 + \sigma^2} e^{\frac{-y^2}{x_2^2 + \sigma^2}} \right) 
        = \frac{-y^2}{\sigma^2} + \log \frac{a_1}{\sigma^2} + \log\left( 1 + \beta e^{\frac{\alpha y^2}{x^2+\sigma^2}}\right),
\end{equation}and by writing the integration interval in the expression~(\ref{eq:Jx}) for $J(x)$ in the form $[0, y_*) \cup [y_*, \infty)$.
Noting that $e^{{-y_*^2}/({x^2 +\sigma^2})} = \beta^{\frac{1}{\alpha}}$,
it is straightforward to compute $J_{21}(x)$, $J_{22}(x)$, $J_{24}(x)$, and $J_{25}(x)$ to obtain
\setlength{\arraycolsep}{0.0em}\begin{eqnarray}
  J_{21}(x) &{}={}& \frac{\beta^{\frac{1}{\alpha}}}{\sigma^2} \left(x^2 + \sigma^2 + y_*^2\right) - 1 - \frac{x^2}{\sigma^2}, \\
  J_{22}(x) &{}={}& \left( 1 - \beta^{\frac{1}{\alpha}}\right) \log \frac{a_1}{\sigma^2}, \\
  J_{24}(x) &{}={}& - \left(\frac{x^2 + \sigma^2 + y_*^2}{x_2^2 + \sigma^2} \right) \beta^{\frac{1}{\alpha}},
\end{eqnarray}\setlength{\arraycolsep}{5pt}and
\setlength{\arraycolsep}{0.0em}\begin{eqnarray}
  J_{25}(x) &{}={}& \left( \log \frac{a_2}{x_2^2 + \sigma^2} \right) \beta^{\frac{1}{\alpha}}.
\end{eqnarray}\setlength{\arraycolsep}{5pt}We now turn our attention to $J_{23}(x)$ and $J_{26}(x)$, which are more delicate to compute. 
To find an expression for $J_{23}(x)$, note that for all $y \in [0, y_*)$ we have $0 < \beta e^{{\alpha y^2}/({x^2+\sigma^2})} < 1$. Therefore, the identity~\cite{table}
\begin{equation}
  \label{eq:seriesexp}
  \log(1+a) = \sum_{k=1}^\infty \frac{(-1)^{k+1}}{k} a^k,
\end{equation}
which is valid for $-1 < a \leq 1$, can be used to obtain
\setlength{\arraycolsep}{0.0em}\begin{eqnarray}
\label{eq:J231} J_{23}(x) &{}={}& \int_0^{y_*} \frac{2y}{x^2 + \sigma^2}e^{\frac{-y^2}{x^2 + \sigma^2}} 
                     \log\left( 1 + \beta e^{\frac{\alpha y^2}{x^2+\sigma^2}}\right) \ud y \\
\label{eq:J232}    & = & \int_0^{y_*} \frac{2y}{x^2 + \sigma^2}e^{\frac{-y^2}{x^2 + \sigma^2}} 
                       \sum_{k=1}^\infty \frac{(-1)^{k+1}}{k} \beta^k e^{\frac{k \alpha y^2}{x^2+\sigma^2}} \ud y \quad \quad \\
\label{eq:J233}    & = & \sum_{k=1}^\infty \int_0^{y_*} \frac{(-1)^{k+1}}{k} \frac{2y \, \beta^k}{x^2 + \sigma^2} e^{\frac{-y^2(1 - k \alpha)}{x^2+\sigma^2}} \ud y \\
\label{eq:J234}    & = & \sum_{k=1}^\infty \frac{(-\beta)^k}{k(k \alpha - 1)} - \beta^{\frac{1}{\alpha}} \sum_{k=1}^\infty \frac{(-1)^k}{k(k \alpha - 1)},
\end{eqnarray}\setlength{\arraycolsep}{5pt}where the order of integration and summation can be inverted in~(\ref{eq:J233}) by virtue of Lebesgue's
dominated convergence theorem~\cite{weir}.
Indeed, defining for $n \in \{1, 2, \ldots \}$
\begin{equation}
  \phi_n(y) \triangleq \frac{2y}{x^2 + \sigma^2}e^{\frac{-y^2}{x^2 + \sigma^2}} \sum_{k=1}^n \frac{(-1)^{k+1}}{k} \beta^k e^{\frac{k \alpha y^2}{x^2+\sigma^2}},
\end{equation}
we see referring to Lemma~\ref{lemma:bound} in Appendix~A that $|\phi_n(y)| \leq \phi_1(y)$ for
all $y \in [0, y_*)$, and since $\int_0^{y_*} \phi_1(y) \ud y < \infty$ the assumptions of Lebesgue's dominated convergence theorem are verified. Evaluating
the sums appearing in~(\ref{eq:J234}) then yields, remembering that $0 < \beta < 1$ and that $\alpha \notin \{1, \frac{1}{2}, \frac{1}{3}, \ldots \}$,
\setlength{\arraycolsep}{0.0em}\begin{eqnarray}
\label{eq:J234b} J_{23}(x) &{}={}& \sum_{k=1}^\infty \frac{(-1)^{k+1}}{k}\beta^k + \sum_{k=1}^\infty \frac{\alpha}{k \alpha - 1}(-\beta)^k 
                                   - \beta^{\frac{1}{\alpha}} \sum_{k=1}^\infty \frac{(-1)^k}{k(k \alpha - 1)} \\
\label{eq:J235}           & = & \log(1 + \beta) - \frac{\alpha \beta}{\alpha - 1}\, \Fto{1, \frac{\alpha-1}{\alpha}}{\frac{2 \alpha - 1}{\alpha}}{-\beta}
                                + \frac{\beta^{\frac{1}{\alpha}}}{\alpha - 1} \, \Ftt{1, 1, \frac{\alpha-1}{\alpha}}{2, \frac{2 \alpha - 1}{\alpha}}{-1},
\end{eqnarray}\setlength{\arraycolsep}{5pt}where 
\begin{equation}
  \label{eq:Fpq}
  \Fpq{\xi_1, \xi_2, \ldots, \xi_p}{\eta_1, \eta_2, \ldots, \eta_q}{z} = 
      \sum_{k=0}^\infty \frac{(\xi_1)_k (\xi_2)_k \ldots (\xi_p)_k}{(\eta_1)_k (\eta_2)_k \ldots (\eta_q)_k} \frac{z^k}{k!}
\end{equation}
denotes the generalised hypergeometric series~\cite{bailey, luke, table}, with $(a)_k \triangleq a(a+1)\cdots(a+k-1)$ the Pochhammer symbol~\cite{table}
and $z \in \mathbb{C}$. In the special case where $p=2$ and $q=1$, the above series reduces to the extensively studied Gauss hypergeometric
series, which is often also denoted $F(\xi_1, \xi_2; \eta_1; z)$\cite{bailey, luke, table}. Although the ${}_3F_2$ appearing in~(\ref{eq:J235})
can be reduced to a ${}_2F_1$ by splitting the second sum appearing in~(\ref{eq:J234}) into partial fractions before summing, we do not do so because
subsequent simplifications in the final expression for $J(x)$ then become less apparent (this remark also applies to~(\ref{eq:J26eval})
below).

Using a similar procedure to evaluate $J_{26}(x)$ yields
\begin{equation}
  \label{eq:J26eval}
  J_{26}(x) = \frac{\beta^{\frac{1}{\alpha}}}{\alpha + 1} \Ftt{1, 1, \frac{\alpha + 1}{\alpha}}{2, \frac{2 \alpha + 1}{\alpha}}{-1}.
\end{equation}

Putting the above results together, we obtain after some elementary manipulations
\setlength{\arraycolsep}{0.0em}\begin{eqnarray}
  J(x) &{}={}& - 1 - \frac{x^2}{\sigma^2} + \log \frac{a_1}{\sigma^2} 
               + \log(1 + \beta) - \frac{\alpha \beta}{\alpha - 1}\, \Fto{1, \frac{\alpha-1}{\alpha}}{\frac{2 \alpha - 1}{\alpha}}{-\beta} \label{eq:JIIcompl} \\
       &     & + \beta^{\frac{1}{\alpha}} \bigg[ \alpha + \frac{1}{\alpha-1} \; 
                                  \Ftt{1, 1, \frac{\alpha-1}{\alpha}}{2, \frac{2 \alpha - 1}{\alpha}}{-1} 
               + \frac{1}{\alpha + 1} \; \Ftt{1, 1, \frac{\alpha + 1}{\alpha}}{2, \frac{2 \alpha + 1}{\alpha}}{-1} \bigg] \nonumber, 
\end{eqnarray}\setlength{\arraycolsep}{5pt}which can be further simplified to obtain
\begin{equation}
  \label{eq:Jxbetasmall}
  J(x) = -1 - \frac{x^2}{\sigma^2} + \log \frac{a_1}{\sigma^2} 
               + \log(1 + \beta) - \frac{\alpha \beta}{\alpha - 1}\, \Fto{1, \frac{\alpha-1}{\alpha}}{\frac{2 \alpha - 1}{\alpha}}{-\beta} 
               + \frac{\pi \beta^{\frac{1}{\alpha}}}{\sin \frac{\pi}{\alpha}}
\end{equation}
as shown in Appendix~B. 

Note that the reason for splitting the integration interval $[0, \infty)$ into $[0, y_*) \cup [y_*, \infty)$ and
evaluating the resulting integrals in different ways is the fact that the series expansion~(\ref{eq:seriesexp}) only is valid for $-1 < a \leq 1$.

\subsubsection{Case III:  $\beta \geq 1$}
\label{sec:caseIII}

We now have
\begin{equation}
  \frac{a_2}{x_2^2 + \sigma^2} e^{\frac{-y^2}{x_2^2 + \sigma^2}} \geq \frac{a_1}{\sigma^2} e^{\frac{-y^2}{\sigma^2}}
\end{equation}
for all $y > 0$, and consequently $J(x)$ can be evaluated by using~(\ref{eq:logid}) together with~(\ref{eq:seriesexp})
over the whole integration range $[0, \infty)$. This yields, using the same technique as in Sec.~\ref{sec:caseII},
$J(x) = J_{31}(x) + J_{32}(x) + J_{33}(x)$, with $J_{31}(x) = J_{11}(x)$ and $J_{32}(x) = J_{12}(x)$ respectively given
in~(\ref{eq:J11}) and~(\ref{eq:J12}), and
\setlength{\arraycolsep}{0.0em}\begin{eqnarray}
  \label{eq:J331}
  J_{33}(x) &{}={}& \int_0^\infty \frac{2y}{x^2 + \sigma^2}e^{\frac{-y^2}{x^2 + \sigma^2}} 
                    \log\left( 1 + \beta^{-1} e^{\frac{-\alpha y^2}{x^2+\sigma^2}}\right) \ud y \quad \\
  \label{eq:J332}
           & = & \log( 1 + \beta^{-1}) - \frac{\alpha \beta^{-1}}{\alpha+1} \; \Fto{1, \frac{\alpha+1}{\alpha}}{\frac{2 \alpha+1}{\alpha}}{-\beta^{-1}}.
\end{eqnarray}\setlength{\arraycolsep}{5pt}$J(x)$ hence reads
\begin{equation}
    \label{eq:Jxbetabig}
    J(x) = -\frac{x^2 + \sigma^2}{x_2^2 + \sigma^2} + \log \frac{a_2}{x_2^2 + \sigma^2} + \log( 1 + \beta^{-1}) 
           -\frac{\alpha \beta^{-1}}{\alpha+1} \; \Fto{1, \frac{\alpha+1}{\alpha}}{\frac{2 \alpha+1}{\alpha}}{-\beta^{-1}}.
\end{equation}
We now give some comments regarding the expressions for $J(x)$ obtained
in~(\ref{eq:Jxbetasmall}) when $0 < \beta < 1$
and $\alpha \notin \{1, \frac{1}{2}, \frac{1}{3}, \ldots \}$, and in~(\ref{eq:Jxbetabig}) when $\beta \geq 1$.

First of all, we recall that in the case when $p = q+1$, the series ${}_pF_q$ defined in~(\ref{eq:Fpq}) converges for $|z| < 1$, also when $z=1$ provided that
\mbox{$\mathrm{Re}\left(\sum{\eta} -\sum \xi\right) > 0$}, and when $z = -1$ provided that \mbox{$\mathrm{Re}\left(\sum{\eta} -\sum \xi\right) > -1$}~\cite{bailey}.
(In our case, when $p = q + 1 = 2$ and $\sum{\eta} -\sum \xi = 0$ the series converges for $z = -1$ as well as for $|z| < 1$,
whereas when $p = q + 1 = 3$ and $\sum{\eta} -\sum \xi = 1$ the series converges for $z = \pm 1$ as well as for $|z| < 1$.) However, ${}_{p+1}F_p$ can be
extended to a single-valued analytic function of $z$ on the domain $\mathbb{C} \backslash (1, \infty)$~\cite{luke, inayat_1, inayat_2},
and it is common practise to use the symbol ${}_{p+1}F_p$ to denote both the resulting generalised
hypergeometric function and the generalised hypergeometric
series on the right hand side of~(\ref{eq:Fpq}) which represents the function inside the unit circle. 

In Appendix~C, we show that the functions $G(z): \mathbb{C} \rightarrow \mathbb{C}$ and $G_2(z): \mathbb{C} \rightarrow \mathbb{C}$, 
which are respectively defined in~(\ref{eq:fzdef}) and~(\ref{eq:f2zdef}), are analytic for all $z \in \mathbb{C} \backslash (-\infty, -1]$. Although the
proof we have is somewhat technical, the consequences are far-reaching. Indeed, if we consider the expression on the right hand side of~(\ref{eq:J331}) as a
function of $z = \beta^{-1}$, then we immediately see that this function is analytic for all $z \in \mathbb{C} \backslash (-\infty, -1]$ by
comparing it to the definition of $G(z)$ in~(\ref{eq:fzdef}). Additionally, the discussion in the previous paragraph shows that the right hand
side of~(\ref{eq:J332}), when considered as a function of $z = \beta^{-1}$ (with $\log(\cdot)$ denoting the principal branch of the
logarithm~\cite{conway}, which is analytic on $\mathbb{C} \backslash(-\infty, 0)$) also is analytic for all $z \in \mathbb{C} \backslash (-\infty, -1]$.
It then follows from the theory of complex analysis~\cite{conway} that equality between the right hand sides of~(\ref{eq:J331}) and~(\ref{eq:J332})
not only holds for $\beta \geq 1$, but actually whenever $\beta^{-1} \in \mathbb{C} \backslash (-\infty, -1]$, and thus in particular for all $\beta > 0$.

When $\alpha \notin \{1, \frac{1}{2}, \frac{1}{3}, \ldots \}$, we can consider the right hand side of~(\ref{eq:J231}) as a function of $z = \beta$,
compare it to the definition of $G_2(z)$ in~(\ref{eq:f2zdef}) to see that it is analytic for all $z \in \mathbb{C} \backslash (-\infty, -1]$, and conclude
as above that equality with the expression in~(\ref{eq:J235}) is assured not only for $0 < \beta < 1$ but
for all $\beta \in \mathbb{C} \backslash (-\infty, -1]$, and thus whenever $\beta > 0$. We can also show, using the same arguments and working with
$G(z)$ in~(\ref{eq:fzdef}), that when $\alpha \notin \{1, \frac{1}{2}, \frac{1}{3}, \ldots \}$, equality between the
right hand sides of~(\ref{eq:J26}) and~(\ref{eq:J26eval}) not only holds
for $0 < \beta < 1$ but actually for all $\beta \in \mathbb{C} \backslash (-\infty, -1]$, and hence whenever $\beta > 0$. We have therefore proved
that, when $\alpha \notin \{1, \frac{1}{2}, \frac{1}{3}, \ldots \}$,~(\ref{eq:Jxbetasmall}) in fact holds for all
$\beta \in \mathbb{C} \backslash (-\infty, -1]$, and as a consequence in particular when $\beta > 0$.

We also would like to discuss in some more detail the restriction $\alpha \notin \{1, \frac{1}{2}, \frac{1}{3}, \ldots \}$ which was made
in Sec.~\ref{sec:caseII}. We see that
\begin{equation}
  \frac{\alpha}{\alpha - 1}\, \Fto{1, \frac{\alpha-1}{\alpha}}{\frac{2 \alpha - 1}{\alpha}}{-\beta}
\end{equation}
is defined neither when $\alpha = 1$ because of the factor $\frac{1}{\alpha-1}$, nor when $\alpha \in \{\frac{1}{2}, \frac{1}{3}, \ldots\}$ 
because in such a case $\frac{2 \alpha - 1}{\alpha}$ is a negative integer or zero, $|\frac{\alpha - 1}{\alpha}| > |\frac{2 \alpha - 1}{\alpha}|$, and 
then the hypergeometric function ${}_2F_1$ itself is not defined~\cite{table, luke}. 
Nonetheless, if we let $d$ belong to the set $\{1, \frac{1}{2}, \frac{1}{3}, \ldots\}$, we still have for all $\beta > 0$ that
\setlength{\arraycolsep}{0.0em}\begin{eqnarray}
                  &   &\lim_{\alpha \rightarrow d} \bigg\{\log(1 + \beta) 
                        - \frac{\alpha \beta}{\alpha - 1}\, \Fto{1, \frac{\alpha-1}{\alpha}}{\frac{2 \alpha - 1}{\alpha}}{-\beta}
                        + \frac{\beta^{\frac{1}{\alpha}}}{\alpha - 1} \, 
                                               \Ftt{1, 1, \frac{\alpha-1}{\alpha}}{2, \frac{2 \alpha - 1}{\alpha}}{-1}\bigg\}\nonumber \\
  \label{eq:lim1} &{}={}&  \lim_{\alpha \rightarrow d} \int_0^{y_*} \frac{2y}{x^2 + \sigma^2}e^{\frac{-y^2}{x^2 + \sigma^2}} 
                     \log\left( 1 + \beta e^{\frac{\alpha y^2}{x^2+\sigma^2}}\right) \, \ud y \\
  \label{eq:lim2} &  =  & \int_0^{y_*} \frac{2y}{x^2 + \sigma^2}e^{\frac{-y^2}{x^2 + \sigma^2}} 
                     \log\left( 1 + \beta e^{\frac{d y^2}{x^2+\sigma^2}}\right) \, \ud y < \infty,
\end{eqnarray}\setlength{\arraycolsep}{5pt}where the exchange of the order of the limit and integration operations again follows from Lebesgue's dominated convergence
theorem (it is easy to verify that the assumptions are met). The last integral can be evaluated using the identity
\setlength{\arraycolsep}{0.0em}\begin{eqnarray}
   \label{eq:indef2}   \int \frac{2y}{a}\;e^{-y^2 \hspace{-1 pt}/a} \log\left( 1 + b \, e^{{d y^2}/{a}}\right) \ud y 
                                        &{}={}& e^{-y^2/a}\sum_{k=0}^{1/d-1} \frac{(-b)^{k}}{\frac{1}{d} - k} e^{k d y^2 / a} \\ 
   & & + \Big((-b)^{\frac{1}{d}} - e^{-y^2 \hspace{-1 pt}/a}\Big)\log\left( 1 + b \, e^{{d y^2}/{a}}\right) - \frac{d (-b)^{\frac{1}{d}}y^2}{a} - d e^{-y^2/a}  \nonumber
\end{eqnarray}\setlength{\arraycolsep}{5pt}which is valid for any $a, b > 0$ and $d \in \{1, \frac{1}{2}, \frac{1}{3}, \ldots\}$
(this can be verified by differentiating the expression on the right-hand side of the equality sign with respect to $y$),
and where the integration constant has been omitted.
For convenience we therefore will consider~(\ref{eq:J235}) and~(\ref{eq:Jxbetasmall}) to be valid also when $\alpha \in \{1, \frac{1}{2}, \frac{1}{3}, \ldots\}$,
it being understood that the value at such points is to be obtained by a limiting process, and that any necessary numerical evaluations can for example
be performed using relations such as~(\ref{eq:indef2}).

We now point the reader's attention to the fact that by setting $x=x_2$ in~(\ref{eq:Jxbetasmall}) and~(\ref{eq:Jxbetabig}), and noting that then
$\alpha = x_2^2 / \sigma^2$ as can be seen from~(\ref{eq:alphabeta}), we obtain the continuation formula
\begin{equation}
  \label{eq:continuation}
  \frac{\alpha \beta}{\alpha - 1} 
  \Fto{1, \frac{\alpha - 1}{\alpha}}{\frac{2 \alpha - 1}{\alpha}}{-\beta} + \alpha -\frac{\pi \beta^{\frac{1}{\alpha}}}{\sin \frac{\pi}{\alpha}}
       = \frac{\alpha \beta^{-1}}{\alpha+1} \Fto{1, \frac{\alpha + 1}{\alpha}}{\frac{2 \alpha + 1}{\alpha}}{-\beta^{-1}}.
\end{equation}
Bearing in mind the remarks in the previous paragraph, this is valid in particular when 
$\alpha, \beta > 0$, the case we are interested in. (Evaluating integrals in more than one way is a technique that has already
been successfully used to derive identities involving the generalised hypergeometric function ${}_3F_2$~\cite{inayat_1, inayat_2}.)

To conclude this section, we note that the hypergeometric functions appearing in~(\ref{eq:Jxbetasmall}) and~(\ref{eq:Jxbetabig}) can also be expressed in terms
of the incomplete beta function $B_z(a,b)$ by virtue of the relation~\cite{table}
\begin{equation}
  \Fto{a, b}{b+1}{z} = b z^{-b} B_z(b, 1-a),
\end{equation}
which is valid whenever the left hand side is defined.

\subsection{Closed-Form Expressions for $I(X; Y)$}

In Sec.~\ref{sec:Jxclosed}, a closed-form expression for $J(x)$ was provided in~(\ref{eq:J11})--(\ref{eq:J13}) for the case
where $\alpha \in \{1, \frac{1}{2}, \frac{1}{3}, \ldots\}$ and $\beta > 0$, and two closed-form expressions for $J(x)$ were provided in~(\ref{eq:Jxbetasmall})
and~(\ref{eq:Jxbetabig}) for the case where $\alpha, \beta > 0$. 

In order to obtain an expression for $I(X;Y)$ which is valid in the general case where $\alpha, \beta > 0$, the two integrals appearing in~(\ref{eq:mutualI}) must
be evaluated using~(\ref{eq:Jxbetasmall}) or~(\ref{eq:Jxbetabig}). There are four different possibilities for how to do this, each with its own advantages and
disadvantages. For example, if one is seeking an analytical expression which does not present indeterminations that must be lifted by a limiting process
for values of $\alpha \in \{1, \frac{1}{2}, \frac{1}{3}, \ldots\}$, one should use~(\ref{eq:Jxbetabig}) to evaluate both integrals in~(\ref{eq:mutualI}).
One might on the other hand be interested in an expression where the series representation~(\ref{eq:Fpq}) of the hypergeometric
function always converges. One then should use~(\ref{eq:Jxbetasmall}) if $\beta < 1$, and~(\ref{eq:Jxbetabig}) if $\beta \geq 1$ to evaluate the integrals
in~(\ref{eq:mutualI}).

Another criterion to decide when to use~(\ref{eq:Jxbetasmall}) and when to use~(\ref{eq:Jxbetabig}) to evaluate the integrals in~(\ref{eq:mutualI})
is the simplicity of the resulting expression for $I(X;Y)$: simple analytical expressions will generally be easier to analyse and provide more
insight than more complex ones. For example, if we choose to evaluate the integrals in~(\ref{eq:mutualI}) by using~(\ref{eq:Jxbetasmall}) and~(\ref{eq:Jxbetabig})
respectively when $k=1$ and $k=2$, we obtain the expression
\setlength{\arraycolsep}{0.0em}\begin{eqnarray}
  I(X;Y)           &{}={}& -(1-a_2) \log(1-a_2) - a_2 \log a_2
                           -(1-a_2) \log(1+\beta) - \frac{a_2 x_2^2}{x_2^2+\sigma^2} \; \Fto{1, -\frac{\sigma^2}{x_2^2}}{-\frac{\sigma^2}{x_2^2}+1}{-\beta} \nonumber \\
  \label{eq:IXYs}  &     & + (1-a_2)\frac{\pi \beta^{1 + \frac{\sigma^2}{x_2^2}}}{\sin \frac{\pi \sigma^2}{x_2^2}} - a_2 \log(1 + \beta^{-1})
                           + \frac{(1 - a_2)\,x_2^2}{\sigma^2}\; \Fto{1, \frac{\sigma^2}{x_2^2} +1}{\frac{\sigma^2}{x_2^2} + 2}{-\beta^{-1}}, 
\end{eqnarray}\setlength{\arraycolsep}{5pt}where we have used the relation $a_1 = 1 - a_2$ to eliminate $a_1$, and one recognises
the entropy $H(X)$ of the channel input $X$, which is distributed according to the law $f_X(x)$ given in~(\ref{eq:inputfx}).
We do not list here all the possible expressions for $I(X;Y)$ which arise from the different ways of evaluating the
integrals in~(\ref{eq:mutualI}), but instead conclude this section with a discussion of the properties of $I(X;Y)$ as $x_2 \rightarrow \infty$
which can be deduced from~(\ref{eq:IXYs}).

If we let $x_2 \rightarrow \infty$ in~(\ref{eq:IXYs}), one can immediately see that 
\begin{equation}
  -(1-a_2) \log(1+\beta) - \frac{a_2 \, x_2^2}{x_2^2+\sigma^2} \; \Fto{1, -\frac{\sigma^2}{x_2^2}}{-\frac{\sigma^2}{x_2^2}+1}{-\beta} \longrightarrow - a_2, \nonumber
\end{equation}
by remembering~(\ref{eq:alphabeta}) and~(\ref{eq:Fpq}). By use of~(\ref{eq:alphabeta}),~(\ref{eq:Fpq}),~(\ref{eq:continuation}),
and L'Hospital's rule, it furthermore follows that
\begin{equation}
  (1-a_2)\frac{\pi \beta^{1 + \frac{\sigma^2}{x_2^2}}}{\sin \frac{\pi \sigma^2}{x_2^2}} \longrightarrow a_2 \nonumber
\end{equation}
and
\begin{equation}
  - a_2 \log(1 + \beta^{-1}) + \frac{(1 - a_2)\,x_2^2}{\sigma^2}\; \Fto{1, \frac{\sigma^2}{x_2^2} +1}{\frac{\sigma^2}{x_2^2} + 2}{-\beta^{-1}}  \longrightarrow 0, \nonumber
\end{equation}
although some effort is required here. As $x_2 \rightarrow \infty$, we thus see that
\begin{equation}
  \label{eq:IXYlim}
  I(X;Y) \longrightarrow -(1-a_2) \log(1-a_2) - a_2 \log a_2 = H(X).
\end{equation}
This can be intuitively understood if one remembers that the mutual information can be written~\cite{cover}
\begin{equation}
  \label{eq:IXYh}
  I(X;Y) = H(X) - H(X|Y),
\end{equation}
with $H(X|Y)$ the conditional entropy of the channel input $X$ given the channel output $Y$. As $x_2 \rightarrow \infty$, the mass points
of the channel input distribution $f_X(x)$ grow more and more separated, and it is to be expected that it will be more and more difficult to make a wrong decision
on the value of the channel input $X$ if the channel output $Y$ were to be known, and hence that $H(X|Y) \rightarrow 0$ as we just proved.

Note in passing that we also have found an exact expression for $H(X|Y)$ as can be seen by comparing~(\ref{eq:IXYs}) with~(\ref{eq:IXYlim}) and~(\ref{eq:IXYh}).

\section{Maximum Attainable Mutual Information and Application to Capacity Calculations}
\label{sec:cap}

We now consider the calculation of the capacity of the channel with transition probability $f_{Y|X}(y|x)$ given in~(\ref{eq:fYX}) when the input $X$ must satisfy
the average power constraint $E[X^2] \leq P$ for a given power budget~$P$. Let the signal to noise ratio (SNR) then be defined by
\begin{equation}
  \mathrm{SNR} \triangleq P /\sigma^2,
\end{equation}
where $\sigma^2$ denotes the noise power (introduced in Sec.~\ref{sec:cm}).
In this case, it has been proved~\cite{abou-faycal} that the capacity-achieving probability density function $f_X(x)$ is discrete
and has a finite number of mass points, one of these mass points being located at the origin ($x=0$).
It has also been empirically found (by means of numerical simulations) that a probability density function $f_X(x)$ with two
mass points achieves capacity for low SNR values, and that the required number of mass points to achieve capacity increases monotonically with the SNR.

Referring to the results from~\cite{abou-faycal}, we see that a distribution with two mass points achieves capacity for SNR values below approximately
$0 \; \mathrm{dB}$, and that the maximum mutual information that can be achieved by using such an input distribution is within $0.02$ nats of the channel capacity
whenever $\mathrm{SNR} \leq 10 \; \mathrm{dB}$.

Following~\cite{abou-faycal}, let $x_1 = 0$ and $x_2^2 = P/a_2$ in $f_X(x)$ (see~(\ref{eq:inputfx})). For any given power budget $P$, the input probability
density function $f_X(x)$
is thus uniquely determined by the value of $a_2$. Furthermore, let\footnote{Note by referring to~(\ref{eq:alphabeta}) and~(\ref{eq:IXYs})
that $I_{a_2}(\SNR)$ only depends on $P$ and~$\sigma^2$ via the ratio $\SNR = P/\sigma^2$.} $I_{a_2}(\mathrm{SNR})$ denote the value taken by the mutual information
$I(X;Y)$ for given values of $a_2$, $P$, and $\sigma^2$ (with $x_2^2 = P/a_2$ in~(\ref{eq:inputfx})), and let
\begin{equation}
  I^*(\mathrm{SNR}) \triangleq \max_{0 \leq a_2 \leq 1} I_{a_2}(\mathrm{SNR})
\end{equation}
and
\begin{equation}
  a_2^*(\mathrm{SNR}) \triangleq \arg \max_{0 \leq a_2 \leq 1} I_{a_2}(\mathrm{SNR}).
\end{equation}
$I^*(\mathrm{SNR})$ hence corresponds to the maximum attainable value of the mutual information $I(X;Y)$ when the input probability
density function $f_X(x)$ has two mass points, and $a_2^*(\mathrm{SNR})$ corresponds to the probability of the nonzero mass point for which $I_{a_2}(\SNR) = I^*(\SNR)$.
The channel capacity $C(\SNR)$ is thus equal to $I^*(\SNR)$ for SNR values below approximately $0 \; \mathrm{dB}$, and exceeds $I^*(\SNR)$ by less than
$0.02$ nats whenever $\SNR \leq 10\; \mathrm{dB}$.

Although $a_2^*(\SNR)$ can be obtained by solving a one-dimensional optimisation problem, it must also satisfy the equation
\begin{equation}
  \label{eq:dIe0}
  \frac{\partial I_{a_2}(\SNR)}{\partial a_2} = 0.
\end{equation}
An expression for $\partial I_{a_2} / \partial a_2$ can be obtained by noting that
\setlength{\arraycolsep}{0.0em}\begin{eqnarray}
\label{eq:dI}
\frac{\ud}{\ud a} \; \Fto{1, h_1(a)}{1 + h_1(a)}{h_2(a)} 
&{}={}& \frac{\ud h_2(a)}{\ud a} \frac{h_1(a)}{1 + h_1(a)} \; \Fto{2, 1 + h_1(a)}{2 + h_1(a)}{h_2(a)} \\
&     & + \; \frac{\ud h_1(a)}{\ud a} \frac{h_2(a)}{(1 + h_1(a))^2} \; \Ftt{2, 1 + h_1(a), 1 + h_1(a)}{2 + h_1(a), 2 + h_1(a)}{h_2(a)} \nonumber
\end{eqnarray}\setlength{\arraycolsep}{5pt}for any given differentiable functions $h_1(\cdot)$ and $h_2(\cdot)$ whenever the left hand-side is defined, and using
this expression together with the expression for $I(X;Y)$ that is obtained by evaluating both integrals in~(\ref{eq:mutualI}) with the help
of~(\ref{eq:Jxbetabig}). (Relation (\ref{eq:dI}) can be established for values of $h_2(\cdot)$ inside the unit circle by using the series representation~(\ref{eq:Fpq}),
and subsequently for all other values of $h_2(\cdot) \in \mathbb{C} \backslash (1, \infty)$ by using analytic continuation arguments.)\footnote{Note that~(\ref{eq:dI})
also could be used to obtain expressions for e.g. $\partial I_{a_2}(\SNR) / \partial \, \SNR$, or for the derivative of $I(X;Y)$ (when the distribution of $X$ is discrete
and has two mass points) with respect to any other parameter of interest.}

It is clear that~(\ref{eq:dIe0}) must be satisfied by at least one value $0 < a_2 < 1$: this can be seen by observing that $I_{a_2}(\SNR) = 0$, 
when $a_2 = 0$ or $a_2 = 1$, and that $I_{a_2}(\SNR) \geq 0$ when $0 < a_2 < 1$. Indeed, $I_{a_2}(\SNR)$ being a continuous and differentiable function
of $a_2$ in this range, it follows that there is at least one value $0 < a_2 < 1$ which satisfies~(\ref{eq:dIe0}). Moreover, numerical experiments also
seem to indicate that the solution of~(\ref{eq:dIe0}) always is unique although we presently do not have a proof of this property.

For SNR values of up to approximately $0 \; \mathrm{dB}$, we have thus found a closed-form expression for the capacity $C(\SNR)$ of the channel with
transition probability~(\ref{eq:fYX}) as a function of a single parameter which can be obtained via numerical root-finding algorithms. 

\begin{figure}
\centering
\includegraphics[width=130 mm]{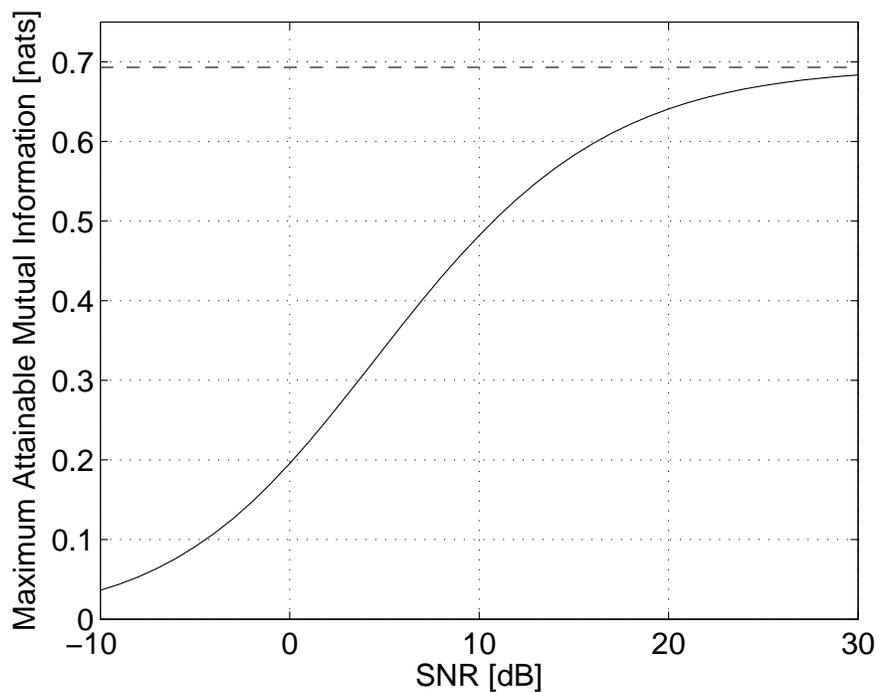}
\caption{Maximum attainable mutual information $I^*_{a_2}(\SNR)$ $[\mathrm{nats}]$ as a function of the SNR
$[\mathrm{dB}]$ when the input probability density function $f_X(x)$ is discrete and restricted to having two mass points. The
constant $\log 2 = 0.693\ldots$ has also been plotted for reference (dashed line).}
\label{fig:IXY_snr}
\end{figure}

\begin{figure}
\centering
\includegraphics[width=130 mm]{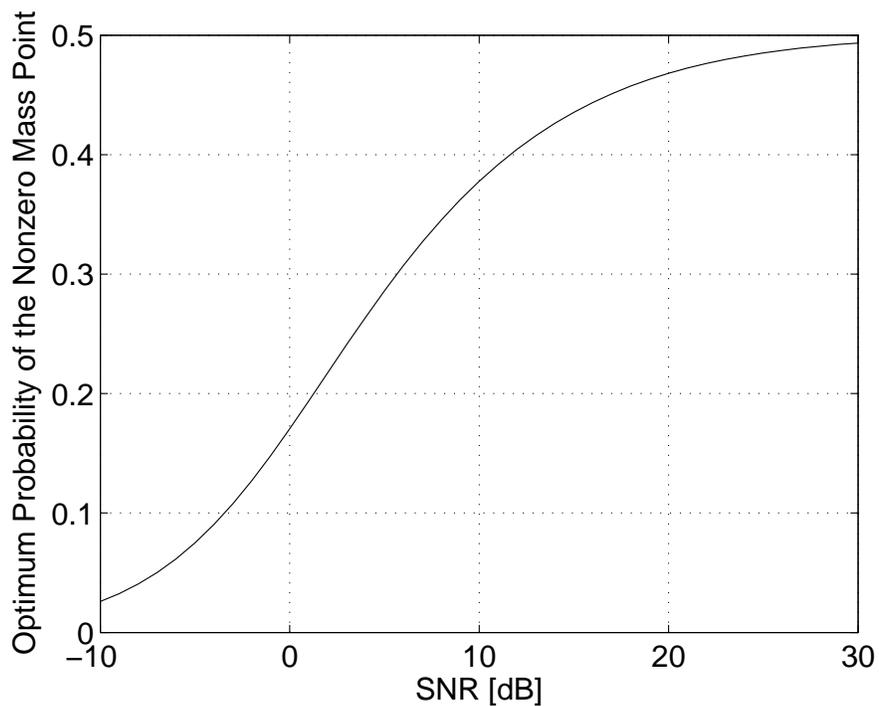}
\caption{Optimal probability $a^*_2(\SNR)$ of the nonzero mass point as a function of the SNR $[\mathrm{dB}]$.}
\label{fig:prob_snr}
\end{figure}

\begin{figure}
\centering
\includegraphics[width=130 mm]{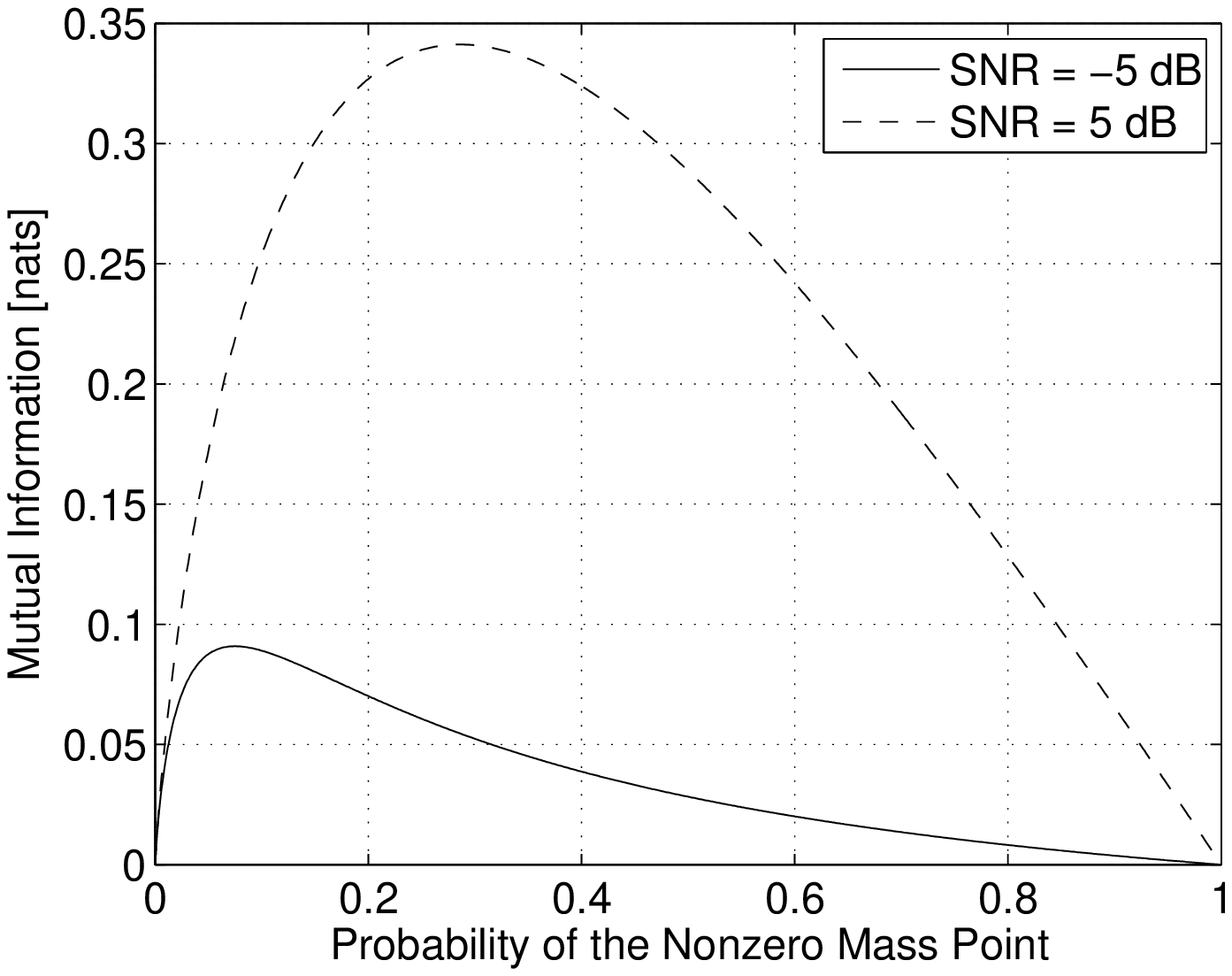}
\caption{Mutual Information $I_{a_2}(\SNR)$ $[\mathrm{nats}]$ as a function of the probability $a_2$ of the
nonzero mass point for $\SNR \in \{-5\, \mathrm{dB}, 5\,\mathrm{dB}\}$.}
\label{fig:IXY_prob}
\end{figure}

Figs.~\ref{fig:IXY_snr} and~\ref{fig:prob_snr} respectively show the maximum attainable mutual information $I^*_{a_2}(\SNR)$ when the input probability density function
$f_X(x)$ is discrete and restricted to having two mass points, and the corresponding optimal probability $a^*_2(\SNR)$ of the nonzero mass point as a function of the SNR.
These figures were obtained by numerically solving~(\ref{eq:dIe0}) for SNR values between $-10\, \mathrm{dB}$ and $\mathrm{30} \, \mathrm{dB}$, which proved to be
by far the simplest method. In Fig.~\ref{fig:IXY_snr} we have also plotted the constant $\log 2 = 0.693\ldots$ for reference, and it can be seen
as expected that  $\lim_{\SNR \rightarrow \infty} I^*_{a_2}(\SNR) = \log 2$, and that $\lim_{\SNR \rightarrow \infty} a_2^*(\SNR) = \textstyle \frac{1}{2}$.

Fig.~\ref{fig:IXY_prob} shows the mutual information $I_{a_2}(\SNR)$ as a function of the probability $a_2$ of the nonzero mass point when
$\SNR \in \{-5\, \mathrm{dB}, 5\,\mathrm{dB}\}$. It is interesting to observe that $I_{a_2}(\SNR)$ ceases to be
concave in $a_2$ for sufficiently low SNR values, and that as $\SNR \rightarrow \infty$, the figure suggests that $I_{a_2}(\SNR)$ indeed approaches the entropy $H(X)$
of the channel input $X$ as also mentioned earlier.\footnote{Note that since $X$ is distributed according to the law $f_X(x)$ given
in~(\ref{eq:inputfx}), $H(X)$ is nothing more than the binary entropy function [nats], which attains its maximum value of $\log 2 = 0.693\ldots$
when $a_1 = a_2 = \frac{1}{2}$.} Note additionally that in both cases shown in the figure the
equation ${\partial I_{a_2}(\SNR)}/{\partial a_2} = 0$ only has one solution $a_2 \in (0, 1)$.

\section{Conclusions}
\label{sec:conclusions}

In this correspondence, we considered the memoryless noncoherent SISO Rayleigh-fading channel. We first derived closed-form expressions
for the mutual information between the output and input signals of this channel when the input magnitude distribution is discrete and restricted to having
two mass points, and subsequently showed how these expressions can be used to derive closed-form expressions for the capacity of this
channel for SNR values below approximately $0 \; \mathrm {dB}$, and a tight capacity lower bound for SNR values between $0 \; \mathrm{dB}$ and
$10 \; \mathrm{dB}$. The expressions for the channel capacity and its lower bound are given as functions of a parameter which can be obtained via
numerical root-finding algorithms.

\appendix

\section*{Appendix A}
\label{sec:Leblemma}

In this appendix, we state and prove a lemma which will be used to show that it is legitimate to exchange the order of integration and
summation in~(\ref{eq:J233}).
\begin{lemma}\label{lemma:bound}
Let $n$ be a positive integer, let $0 \leq q \leq 1$, and let $T_n \triangleq \sum_{k=1}^n \frac{(-1)^{k+1}}{k} q^k$. Then 
$0 \leq T_n \leq q$ for all $n$.
\end{lemma}
\begin{proof} Let $n$ and $p$ be positive integers. The result is trivial if $q = 0$. We thus assume $0 < q \leq 1$ in the remainder of this proof. 
For all such $q$, we always have $T_1 = q$, and $T_{2n} < T_{2n+1}$. The lemma will thus be proved
if we show that $0 \leq T_{2n}$ and $T_{2n+1} \leq q$ for any positive integer $n$. Now, noting that
\begin{equation}
  c_p \triangleq \frac{q^p}{p} - \frac{q^{p+1}}{p+1} = \frac{q^p}{p} - \frac{q^p}{p} \frac{qp}{p+1} > 0,
\end{equation}
we see that $T_{2n} = \sum_{k=0}^{n-1}c_{2k+1} > 0$ and that $T_{2n+1} = q - \sum_{k=1}^n c_{2k} < q$.
\end{proof}

\section*{Appendix B}
\label{sec:simpli}

In this appendix, we show how the expression for $J(x)$ given in~(\ref{eq:JIIcompl}) can be reduced to the simpler
equivalent expression given in~(\ref{eq:Jxbetasmall}).
Let\setlength{\arraycolsep}{0.0em}\begin{eqnarray}
  S&{}\triangleq{}& \alpha + \frac{1}{\alpha-1} \Ftt{1, 1, \frac{\alpha-1}{\alpha}}{2, \frac{2 \alpha - 1}{\alpha}}{-1} 
                  + \frac{1}{\alpha + 1} \Ftt{1, 1, \frac{\alpha + 1}{\alpha}}{2, \frac{2 \alpha + 1}{\alpha}}{-1} \nonumber \\
   &  =  & \alpha + \sum_{k=1}^\infty \frac{(-1)^{k+1}}{k(k \alpha - 1)} + \sum_{k=1}^\infty \frac{(-1)^{k+1}}{k(k \alpha + 1)} \nonumber \\
   &  =  & \alpha + \sum_{k=0}^\infty \frac{1}{(2k+1)((2k+1)\alpha - 1)} - \sum_{k=1}^\infty \frac{1}{2k(2k \alpha - 1)} \nonumber \\
   &     & + \sum_{k=0}^\infty \frac{1}{(2k+1)((2k+1)\alpha + 1)} - \sum_{k=1}^\infty \frac{1}{2k(2k \alpha + 1)} \nonumber \\
   &  \triangleq  & \alpha + S_1 - S_2 + S_3 - S_4.  \label{eq:S}
\end{eqnarray}\setlength{\arraycolsep}{5pt}with obvious definitions for $S_1$, $S_2$, $S_3$, and $S_4$,
and the assumptions of Sec~\ref{sec:caseII} still applying. These sums can be evaluated in terms
of the psi function $\psi(\cdot)$~\cite{table}, which admits the series representation
\setlength{\arraycolsep}{0.0em}\begin{eqnarray}
  \psi(q+1) &{}={}&  -\gamma + \sum_{k=0}^\infty \frac{q}{(k+1)(k+1+q)} \label{eq:psiseries1} \\
            &  =  & -\gamma + \sum_{k=1}^\infty \frac{q}{k(k+q)}, \label{eq:psiseries2}
\end{eqnarray}\setlength{\arraycolsep}{5pt}with $\gamma = 0.577\ldots$ denoting Euler's constant~\cite{table},
as can easily be deduced from~\cite[8.362 (1)]{table}. This yields
\setlength{\arraycolsep}{0.0em}\begin{eqnarray}
  S_1 &{}={}& - \frac{\gamma}{2} -\log 2 -\frac{1}{2}\psi\Big(\frac{\alpha-1}{2 \alpha}\Big), \label{eq:S_1} \\
  S_2 &  =  & - \frac{\gamma}{2} -\frac{1}{2}\psi\Big(\frac{2 \alpha-1}{2 \alpha}\Big), \label{eq:S_2} \\
  S_3 &  =  & \frac{\gamma}{2}  + \log 2 + \frac{1}{2}\psi\Big(\frac{\alpha+1}{2 \alpha}\Big), \label{eq:S_3}
\end{eqnarray}\setlength{\arraycolsep}{5pt}and
\begin{equation}
  S_4 = \frac{\gamma}{2}  + \alpha + \frac{1}{2}\psi\Big(\frac{1}{2 \alpha}\Big). \label{eq:S_4}  
\end{equation}
Here,~(\ref{eq:S_1}) can be verified by setting $q = -\frac{1}{2\alpha}-\frac{1}{2}$ in~(\ref{eq:psiseries1}) to evaluate
$S_1 + \frac{\gamma}{2} + \frac{1}{2}\psi(\frac{\alpha-1}{2 \alpha})$, and noting that
\setlength{\arraycolsep}{0.0em}\begin{eqnarray}
  \log 2 &{}={}& \textstyle (1 - \frac{1}{2}) + (\frac{1}{3} - \frac{1}{4}) + (\frac{1}{5} - \frac{1}{6}) + \cdots \label{eq:logseries1}\\
         &  =  & \sum_{k=1}^\infty \frac{1}{(2k+1)(2k+2)}, \label{eq:logseries2}
\end{eqnarray}\setlength{\arraycolsep}{5pt}with~(\ref{eq:logseries1}) following from~(\ref{eq:seriesexp}) at $a=1$; (\ref{eq:S_2}) follows directly
by setting $q = -\frac{1}{2 \alpha}$ in~(\ref{eq:psiseries2}); (\ref{eq:S_3}) can be obtained by using~(\ref{eq:psiseries1}) with $q = \frac{1}{2\alpha} - \frac{1}{2}$
to evaluate $\frac{\gamma}{2} + \frac{1}{2}\psi(\frac{\alpha+1}{2\alpha}) - S_3$ and remembering~(\ref{eq:logseries2}); and~(\ref{eq:S_4}) can be proved
by setting $q = \frac{1}{2 \alpha} - 1$ in the identity
\begin{equation}
  \psi(q+1) = -\gamma + \frac{q}{q+1} + \sum_{k=1}^\infty \frac{q}{(k+1)(k+1+q)},
\end{equation}
which can easily be obtained starting from~(\ref{eq:psiseries1}), to find an expression for
$\frac{\gamma}{2} + \frac{1}{2} \psi(\frac{1}{2 \alpha}) + \alpha - S_4$ and observing that
\setlength{\arraycolsep}{0.0em}\begin{eqnarray}
  1 &{}={}& \textstyle (1 - \frac{1}{2}) + (\frac{1}{2} - \frac{1}{3}) + (\frac{1}{3} - \frac{1}{4}) + \cdots = \displaystyle \sum_{k=1}^\infty \frac{1}{k(k+1)}.  \qquad
\end{eqnarray}\setlength{\arraycolsep}{5pt}Combining~(\ref{eq:S}) with (\ref{eq:S_1})--(\ref{eq:S_4}) yields
\setlength{\arraycolsep}{0.0em}\begin{eqnarray}
  S &{}={}& \textstyle \frac{1}{2} \left[   \psi(1 - \frac{1}{2\alpha}) - \psi(\frac{1}{2\alpha}) 
                               + \psi(\frac{1}{2} + \frac{1}{2\alpha}) - \psi(\frac{1}{2} - \frac{1}{2\alpha})\right] \qquad \\
    & =  & \textstyle \frac{1}{2} \left[ \pi \cot \frac{\pi}{2 \alpha} + \pi \tan \frac{\pi}{2 \alpha} \right] \label{eq:Sint} \\
    & = & \frac{\pi}{\sin \frac{\pi}{\alpha}}, \label{eq:Sfi}
\end{eqnarray}\setlength{\arraycolsep}{5pt}where~(\ref{eq:Sint}) follows from the relations~\cite{table}
\setlength{\arraycolsep}{0.0em}\begin{eqnarray}
  \psi(1-q) &{}={}& \psi(q) + \pi \cot \pi q \\
  \textstyle \psi(\frac{1}{2}+q) &{}={}& \textstyle \psi(\frac{1}{2} - q) + \pi \tan \pi q.
\end{eqnarray}\setlength{\arraycolsep}{5pt}The simplification in~(\ref{eq:JIIcompl}) now follows immediately from~(\ref{eq:S}) and~(\ref{eq:Sfi}).

\section*{Appendix C}
\label{sec:analytic}

In this appendix, we show that the functions $G(z)$ and $G_2(z)$, defined respectively in~(\ref{eq:fzdef}) and~(\ref{eq:f2zdef}), are analytic for
all $z \in \mathbb{C} \backslash (-\infty, -1]$.
This fact is used in Sec.~\ref{sec:Jxclosed} to prove additional properties of the expressions for $J(x)$
given in~(\ref{eq:Jxbetasmall}) and~(\ref{eq:Jxbetabig}).

Let us consider the function $G: \mathbb{C} \rightarrow \mathbb{C}$ defined by
\begin{equation}
  \label{eq:fzdef}
  G(z) = \int_c^\infty {2 a y} \, e^{-a y^2} \log \Big( 1 + z \, e^{-b y^2}\Big) \, \ud y,
\end{equation}
with $a, b$ positive real constants, $c$ a nonnegative real constant, and where $\log(\cdot)$ denotes the principal branch of the
logarithm~\cite{conway}. 

To prove that $G(z)$
is analytic for all $z \in \mathbb{C} \backslash (-\infty, -1]$, consider the function $g: \mathbb{C} \times \mathbb{R}^+  \rightarrow \mathbb{C}$ defined by
\begin{equation}
  \label{eq:gzy}
  g(z, y) = {2 a y} \, e^{-a y^2} \log \Big( 1 + z \, e^{-b y^2}\Big).
\end{equation}
For any fixed $y$ it is clear that $g(z, y)$, when considered as a function of $z$ only, is analytic on
the domain $z \in \mathbb{C} \backslash (-\infty, -1]$ since the principal branch of the logarithm is analytic on $\mathbb{C} \backslash (-\infty, 0]$.
It hence satisfies the Cauchy-Riemann equations (which are a necessary and sufficient condition for a complex function to be analytic)~\cite{conway}
\begin{equation}
  \frac{\partial u_g}{\partial z_r} = \frac{\partial v_g}{\partial z_i} \quad \mbox{and} \quad \frac{\partial u_g}{\partial z_i} = -\frac{\partial v_g}{\partial z_r}
   \label{eq:CauchyRiemanng}
\end{equation}
for all $z \in \mathbb{C} \backslash (-\infty, -1]$, where $u_g(z_r, z_i, y) = \mathrm{Re} \, g(z_r + i z_i, y)$ and $v_g(z_r, z_i, y) = \mathrm{Im} \, g(z_r + i z_i, y)$.

In order to show that $G(z)$ is analytic on $\mathbb{C} \backslash (-\infty, -1]$, we will show that the Cauchy-Riemann equations
\begin{equation}
  \frac{\partial u_G}{\partial z_r} = \frac{\partial v_G}{\partial z_i} \quad \mbox{and} \quad \frac{\partial u_G}{\partial z_i} = -\frac{\partial v_G}{\partial z_r}
     \label{eq:CauchyRiemannf}
\end{equation}
are satisfied for all $z \in \mathbb{C} \backslash (-\infty, -1]$, where $u_G(z_r, z_i) = \mathrm{Re} \; G(z_r + i z_i)$ and $v_G(z_r, z_i) = \mathrm{Im} \; G(z_r + i z_i)$.
(We will sometimes write $u_G(z)$ instead of $u_G(z_r, z_i)$, $v_G(z)$ instead of $v_G(z_r, z_i)$, $u_g(z, y)$ instead of $u_g(z_r, z_i, y)$, 
and $v_g(z, y)$ instead of $v_g(z_r, z_i, y)$ for convenience.)
Since obviously
\begin{equation}
  u_G(z) = \int_c^\infty u_g(z, y) \, \ud y \quad \mbox{and} \quad v_G(z) = \int_c^\infty v_g(z, y) \, \ud y,
\end{equation}
it is sufficient to show that
\begin{equation}
  \frac{\partial}{\partial z_r} \int_c^\infty u_g(z, y) \, \ud y = \int_c^\infty \frac{\partial u_g(z, y)}{\partial z_r} \, \ud y, \label{eq:CR1}
\end{equation}
together with similar conditions for 
\begin{equation}
  \frac{\partial}{\partial z_i} \int_c^\infty u_g(z, y) \, \ud y, \quad \frac{\partial}{\partial z_r} \int_c^\infty v_g(z, y) \, \ud y, \quad
  \frac{\partial}{\partial z_i} \int_c^\infty v_g(z, y) \, \ud y \label{eq:CRR}
\end{equation}
by virtue of~(\ref{eq:CauchyRiemanng}). In order to do so, we will use the following result~\cite{weir}:

\begin{theorem} \label{theorem:dui}
Let $(\mathcal{Y}, \mathcal{A}, \mu)$ (with $\mathcal{Y}$ a set, $\mathcal{A}$ a $\sigma$-algebra on $\mathcal{Y}$,
and $\mu$ a measure on $\mathcal{A}$) be a measure space, let
$(b_0, b_1)$ be a (possibly infinite) interval of $\mathbb{R}$, and let $h:(b_0, b_1) \times \mathcal{Y} \rightarrow \mathbb{R} \cup \infty$.
For all $x \in (b_0, b_1)$, let
\begin{equation}
  \mathcal{J}(x) = \int_\mathcal{Y} h(x, y) \, \ud \mu
\end{equation}
which is assumed to be always finite. Let $\mathcal{V}$ be a neighbourhood of $x^* \in (b_0, b_1)$ such that:
\begin{enumerate}
\item[(i)] For almost all $y \in \mathcal{Y}$, $h(x, y)$ is continuously differentiable with respect to $x$ on $\mathcal{V}$.
\item[(ii)] There exists an integrable function $\mathcal{H}:\mathcal{Y} \rightarrow \mathbb{R} \cup \infty$ such that for all $x \in \mathcal{V}$
  \begin{equation}
    \left| \frac{\partial h}{\partial x}(x, y) \right| \leq \mathcal{H}(y) \;\; \mbox{almost everywhere.}
  \end{equation}
\end{enumerate}
Then $\mathcal{J}(x)$ is differentiable at $x^*$, and
\begin{equation}
  \frac{\ud \mathcal{J}(x^*)}{\ud x} = \int_\mathcal{Y} \frac{\partial h}{\partial x}(x^*, y) \, \ud \mu.
\end{equation}
\end{theorem}
\begin{proof}
  See standard texts on Lebesgue integration such as~\cite{weir}.
\end{proof}
In our case, $\mathcal{Y}$ is the interval $[c, \infty)$, $\mathcal{A}$ is the Borel $\sigma$-algebra on
$\mathcal{Y}$, and $\mu$ is the standard Lebesgue measure on $\mathcal{A}$. The function
$h(z_r, y) = u_g(z_r, z_i, y)$ with $z_i$ fixed when proving~(\ref{eq:CR1}), and similar definitions when proving the conditions in~(\ref{eq:CRR}).

Now, for the $g(z, y)$ defined in~(\ref{eq:gzy}), we have
\setlength{\arraycolsep}{0.0em}\begin{eqnarray}
  u_g(z_r, z_i, y) &{}={}& 2 a y e^{-ay^2} \log \left| 1 +(z_r + i z_i) e^{by^2} \right| \nonumber \\
                  &{}={}& a y e^{-ay^2} \log \left( \big(1+z_r e^{-by^2}\big)^2 + \big( z_i e^{-by^2}\big)^2 \right), \qquad 
\end{eqnarray}\setlength{\arraycolsep}{5pt}and hence
\begin{equation}
  \label{eq:partialug}
  \frac{\partial u_g}{\partial z_r} = \frac{2ay e^{-(a+b)y^2} \big(1+z_r e^{-by^2}\big)}
                                           {\big(1+z_r e^{-by^2}\big)^2 + \big( z_i e^{-by^2}\big)^2}  = \frac{\partial v_g}{\partial z_i}.
\end{equation}

In the sequel, we will need to upper bound the absolute value of~(\ref{eq:partialug}) in different settings. This will always be accomplished
by obtaining an upper bound $M < \infty$ for $\big|1+z_r e^{-by^2}\big|$ and a lower bound $m > 0$ for $\big|\big(1+z_r e^{-by^2}\big)^2 + \big( z_i e^{-by^2}\big)^2\big|$.

Let $z_r^*$ and $z_i$ be such that $z_r^*+iz_i \in \mathbb{C} \backslash (-\infty, -1]$, and let the neighbourhood $\mathcal{V}$ of $z_r^*$ be defined by
\begin{equation}
  \mathcal{V} = \left\{\begin{array}{rl} (z^*_r - \varepsilon, z^*_r + \varepsilon) & \mbox{if } z_i = 0 \mbox{ and } z_r^* \in (-1, \infty) \\
                               (z^*_r-1, z^*_r+1) & \mbox{if } z_i \neq 0,
             \end{array}\right.
\end{equation}
with $\varepsilon \triangleq \frac{z^*_r+1}{2} > 0$ when $z_i = 0$ and $z_r^* \in (-1, \infty)$. Then,
\begin{equation}
  \big| 1 + z_r e^{-by^2} \big| \leq 1 + |z_r| \leq 1 + |z_r^*| + \max\{1, \varepsilon\} \triangleq M,
\end{equation}
for any $z_r \in \mathcal{V}$, any given $z_i$, and any $y \in [c, \infty)$; whereas if $z_i = 0$, 
\setlength{\arraycolsep}{0.0em}\begin{eqnarray}
   {\big(1+z_r e^{-by^2}\big)^2 + \big( z_i e^{-by^2}\big)^2} &{}={}& \big(1+z_r e^{-by^2}\big)^2 \nonumber \\
                                                           & \geq & \textstyle \big(1 + \frac{z_r^* -1}{2} e^{-by^2} \big)^2 \nonumber \\
                                                           & \geq & \textstyle \min \left\{1, \big(1 + \frac{z_r^* -1}{2}\big)^2 \right\} \triangleq m_1 \qquad
\end{eqnarray}\setlength{\arraycolsep}{5pt}for any $z_r \in \mathcal{V}$ and any $y \in [c, \infty)$, and if $z_i \neq 0$ we have for any $z_r \in \mathcal{V}$
and any $y \in [c, \infty)$
\setlength{\arraycolsep}{0.0em}\begin{eqnarray}
  \big(1+z_r e^{-by^2}\big)^2 &{}+{}&\big( z_i e^{-by^2}\big)^2 \nonumber\\
      &{}\geq{}& \textstyle \min \left\{ \big(1+z_r e^{-bc^2}\big)^2 + \big( z_i e^{-bc^2}\big)^2, 1, \frac{z_i^2}{z_r^2 + z_i^2}\right\} \nonumber \\
      &{}\geq{}& \textstyle \min \left\{ \big( z_i e^{-bc^2}\big)^2, 1, \frac{z_i^2}{\max\{(z_r^*-1)^2, (z_r^*+1)^2\} + z_i^2}\right\} \nonumber \\
      & \triangleq &  m_2,
\end{eqnarray}\setlength{\arraycolsep}{5pt}as one can see by noting that 
\begin{equation}
  \inf_{y \in [c, \infty)} \; \big(1+z_r e^{-by^2}\big)^2+\big( z_i e^{-by^2}\big)^2
\end{equation}
must either be attained when $y=c$, when $y \rightarrow \infty$, or when
\begin{equation}
  \frac{\partial}{\partial y} \left\{ \big(1+z_r e^{-by^2}\big)^2+\big( z_i e^{-by^2}\big)^2 \right\} = 0.
\end{equation}
Consequently, if we set $m \triangleq \min\{m_1, m_2\} > 0$, we see that for all $z_r \in \mathcal{V}$ and any $y \in [c, \infty)$,
\begin{equation}
  \left|\frac{\partial u_g}{\partial z_r} (z_r, z_i, y)\right| \leq \frac{2 M a y \, e^{-(a+b)y^2}}{m},
\end{equation}
which is integrable on $[c, \infty)$ for any $a, b, m, M>0$ and any $c \geq 0$. Noting in addition that for any $z_r \in \mathcal{V}$ and any $z_i$,
$\frac{\partial u_g}{\partial z_r}$ given in~(\ref{eq:partialug}) is continuous in $z_r$, we have proved that~(\ref{eq:CR1}) holds for all
$z = z_r^* + iz_i \in \mathbb{C}\backslash(-\infty, -1]$ by virtue of Theorem~\ref{theorem:dui}.

Similarly, let $z_r$ and $z_i^*$ be such that $z_r+iz_i^* \in \mathbb{C} \backslash (-\infty, -1]$, and let the neighbourhood $\mathcal{V}$ of $z_i^*$ be defined by
\begin{equation}
  \mathcal{V} = \left\{\begin{array}{rl} (z^*_i - 1, z^*_i + 1) & \mbox{if } z_r \in (-1, \infty) \\
                               (z^*_i-\varepsilon, z^*_i+\varepsilon) & \mbox{if } z_r \in (-\infty, -1] \mbox{ and } z^*_i \neq 0,
             \end{array}\right.
\end{equation}
with $\varepsilon = \frac{|z_i^*|}{2} > 0$ when $z^*_i \neq 0$. Then,
\begin{equation}
  \big| 1 + z_r e^{-by^2} \big| \leq 1 + |z_r| \triangleq M',
\end{equation}
for any $z_i \in \mathcal{V}$ and any $y \in [c, \infty)$; whereas if $z_r \in (-1, \infty)$,
\setlength{\arraycolsep}{0.0em}\begin{eqnarray}
  \big(1+z_r e^{-by^2}\big)^2 +\big( z_i e^{-by^2}\big)^2 &{}\geq{} & \big(1+z_r e^{-by^2}\big)^2 \nonumber \\
      &{}\geq{} & \min \left\{ 1, \big(1+z_r\big)^2 \right\} \triangleq m_1', \qquad
\end{eqnarray}\setlength{\arraycolsep}{5pt}for all $z_i \in \mathcal{V}$ and any $y \in [c, \infty)$, and if $z_i^* \neq 0$
\setlength{\arraycolsep}{0.0em}\begin{eqnarray}
  \big(1&{}+{}&z_r e^{-by^2}\big)^2 +\big( z_i e^{-by^2}\big)^2 \nonumber\\
      &{}\geq{}& \textstyle \min \left\{ \big(1+z_r e^{-bc^2}\big)^2 + \big( z_i e^{-bc^2}\big)^2, 1, \frac{z_i^2}{z_r^2 + z_i^2}\right\} \nonumber \\
      &{}\geq{}& \textstyle \min \left\{ \big( (z_i^* - \varepsilon) e^{-bc^2}\big)^2, 1, 
                 \frac{\min\{(z_i^*-\varepsilon)^2, (z_i^*+\varepsilon)^2\}}{z_r^2 + \min\{(z_i^*-\varepsilon)^2, (z_i^*+\varepsilon)^2\}}\right\} \nonumber \\
      & \triangleq &  m_2',
\end{eqnarray}\setlength{\arraycolsep}{5pt}for all $z_i \in \mathcal{V}$ and any $y \in [c, \infty)$.
Consequently, if we set $m' \triangleq \min\{m_1', m_2'\} > 0$, we see that for all $z_i \in \mathcal{V}$ and any $y \in [c, \infty)$,
\begin{equation}
  \left|\frac{\partial v_g}{\partial z_i} (z_r, z_i, y)\right| \leq \frac{2 M' a y \, e^{-(a+b)y^2}}{m'},
\end{equation}
which is integrable on $[c, \infty)$ for any $a, b, m', M'>0$ and any $c \geq 0$. Noting in addition that for any $z_i \in \mathcal{V}$ and any $z_r$,
$\frac{\partial v_g}{\partial z_i}$ given in~(\ref{eq:partialug}) is continuous in $z_i$, we have proved that the order of integration
and derivation can be exchanged in the last expression appearing in~(\ref{eq:CRR}) for all $z = z_r + i z_i^* \in \mathbb{C} \backslash (-\infty, -1]$.

Noting now that
\begin{equation}
  \frac{\partial u_g}{\partial z_i} = \frac{2ay e^{-(a+b)y^2} \big(z_i e^{-by^2}\big)}
                                           {\big(1+z_r e^{-by^2}\big)^2 + \big( z_i e^{-by^2}\big)^2}  = -\frac{\partial v_g}{\partial z_r},
\end{equation}
it is easy to transpose the above arguments to prove that exchanging the order of differentiation and integration is legitimate 
for all $z \in \mathbb{C} \backslash (-\infty, -1]$ also in the case of
the first and second expressions appearing in~(\ref{eq:CRR}), and we have thus established the analytic character of the function $G(z)$ defined
in~(\ref{eq:fzdef}) on the domain $\mathbb{C} \backslash (-\infty, -1]$.

Let us now consider the function $G_2: \mathbb{C} \rightarrow \mathbb{C}$ defined by
\begin{equation}
  \label{eq:f2zdef}
  G_2(z) = \int_0^c {2 a y} \, e^{-a y^2} \log \Big( 1 + z \, e^{-b y^2}\Big) \, \ud y,
\end{equation}
with $a$ a positive real constant, $b \in \mathbb{R}$, and $c$ a nonnegative real constant. The arguments from this appendix can be used
verbatim to prove that $G_2(z)$ also is analytic on $\mathbb{C} \backslash (-\infty, -1]$ -- and the proof can even be simplified due to the
fact that the integration interval $[0, c)$ now being finite, it is sufficient that the corresponding functions $u_{g_2}(z_r, z_i, y)$ and $v_{g_2}(z_r, z_i, y)$
be differentiable functions of $z_r$ and $z_i$ on $\mathbb{C} \backslash (-\infty, -1]$ for all $y \in [0, c)$ to guarantee that the order in which 
integration and differentiation are performed in~(\ref{eq:CR1}) and in the expressions given in~(\ref{eq:CRR}) (with $u_g(z,y)$ and $v_g(z,y)$
respectively replaced by $u_{g_2}(z,y)$ and $v_{g_2}(z,y)$, and the integration interval $[c, \infty)$ replaced by $[0,c)$)
can be exchanged~\cite[Sec.~12.211]{table}.

\bibliographystyle{unsrt}
\bibliography{mybib}

\end{document}